\documentstyle[aps,multicol]{revtex}
\renewcommand{\narrowtext}{\begin{multicols}{2} \global\columnwidth20.5pc}
\renewcommand{\widetext}{\end{multicols} \global\columnwidth42.5pc}

\input{epsf.tex}
\def\inseps#1#2{\def\epsfsize##1##2{#2##1} \centerline{\epsfbox{#1}}}

\begin{document}
\draft

\title{Solitary Waves of Planar Ferromagnets and the Breakdown of the
Spin-Polarized Quantum Hall Effect}

\author{N.R. Cooper}
\address{Institut Laue-Langevin, Avenue des Martyrs, B.P. 156, 38042
Grenoble, France\\
and T.C.M. Group, Cavendish Laboratory, Madingley Road, Cambridge CB3 0HE,
United Kingdom.}
\date{28 December, 1997}

\maketitle

\begin{abstract}

A branch of uniformly-propagating solitary waves of planar
ferromagnets is identified.  The energy dispersion and structures of
the solitary waves are determined for an isotropic ferromagnet as
functions of a conserved momentum.  With increasing momentum, their
structure undergoes a transition from a form ressembling a droplet of
spin-waves to a Skyrmion/anti-Skyrmion pair.  An instability to the
formation of these solitary waves is shown to provide a mechanism for
the electric field-induced breakdown of the spin-polarized quantum
Hall effect.

\end{abstract}



\narrowtext

Continuum descriptions of the static and dynamic properties of
ferromagnets give rise to many forms of non-linear
structure\cite{kosevich}.  These structures depend strongly on the
symmetries and dimensionality of the ferromagnet.  In two dimensions,
the existence of a topological classification of spin-configurations
\cite{rajaraman} has important consequences.  In particular, the
topological index plays a key role in the construction of a sequence
of static ``Skyrmion'' solitons of an isotropic
ferromagnet\cite{belavinpolyakov}.  These solitons have been of recent
interest in quantum Hall systems, for which the low-energy charged
excitations can be described as Skyrmions \cite{sondhi}.

The topological density is of particular importance for the dynamical
properties of planar ferromagnets.  It has been established that the
dynamics are contrained not only by the existence of the topological
index, but also by conservation laws based on moments of the
topological density\cite{paptom}.  Here, we use the properties of a
conserved momentum, defined in terms of the dipole moment of the
topological density, to identify a branch of uniformly-propagating
solitary waves of planar ferromagnets.  At large momentum the solitary
waves appear as Skyrmion/anti-Skyrmion pairs, which move at a constant
velocity perpendicular to the line separating their centres.  To the
best of our knowledge, these are the first propagating solitary waves
of a planar ferromagnet to be identified.  We further show that an
instability to the creation of these solitary waves leads to a
breakdown of the spin-polarized quantum Hall effect, akin to the
destruction of superfluid flow by vortex-ring
nucleation\cite{langerreppy}.

We start from the dissipationless Landau-Lifshitz equation, which
provides a semiclassical description of 
the dynamics of the local magnetization of a ferromagnet, as
represented by the 3-component unit vector
$n_i(\bbox{r},t)$\cite{landaulifshitz}
\begin{equation}
\label{eq:ll}
\rho J \frac{\partial n_i}{\partial t} =  - \epsilon_{ijk}
n_j\frac{\delta E}{\delta n_k} .
\end{equation}
$\rho$ is the density of magnetic moments, each of angular
momentum $J$, and $E(n_i, \nabla_\alpha n_j)$ is an energy functional
of the magnetization and its spatial derivatives.  (Summation
convention is assumed throughout, the labels $i,j,k$ running over the
three spin-components and $\alpha,\beta$ over the two spatial
dimensions.)

The total energy, $E$, and the modulus, $n_i^2$, are conserved
by Eq.(\ref{eq:ll}). 
There are several other conserved quantities that are
important in what follows: the topological index, $Q$, the linear
momentum, $P_\alpha$, and the total number of spin-reversals, $N$,
defined by\cite{paptom}
\begin{eqnarray}
Q & \equiv & \int d^2\bbox{r}\; q(\bbox{r}) , \\
\label{eq:mtm}
P_\alpha & \equiv & 4\pi \rho J \epsilon_{\alpha\beta} \int
d^2\bbox{r}\; r_\beta q(\bbox{r}) , \\
N & \equiv &  \rho J/\hbar \int d^2\bbox{r}\; (1-n_z) ,
\label{eq:num}
\end{eqnarray}
where $q(\bbox{r})\equiv
\frac{1}{8\pi}\epsilon_{ijk}\epsilon_{\alpha\beta} n_i\nabla_\alpha
n_j\nabla_\beta n_k$ is the topological density.  

We will focus on configurations for which the magnetization tends to
$n_z=1$ at spatial infinity. In this case, the topological index $Q$
is an integer\cite{rajaraman,belavinpolyakov}, that is conserved under
smooth deformations of $n_i(\bbox{r})$.  $Q$ is therefore a constant
of any regular dynamics.  The conservation of $P_\alpha$ and $N$ under
Eq.~(\ref{eq:ll}) requires certain restrictions on the terms in the
energy functional. It is sufficient that the energy functional
describe a translationally-invariant ferromagnet, with at most uniaxial
anisotropy (chosen along the $\hat{z}$-axis), and that magnetic
dipolar interactions can be neglected\cite{paptom}.  These conditions
are met by the energy functionals considered below.

Essential to our work is the definition of linear momentum
(\ref{eq:mtm}) in terms of the dipole moment of the topological
density. This form avoids ambiguities inherent in other
definitions\cite{kosevich} (see Ref.\onlinecite{paptom}), that would
prove important for the configurations studied below.  We use the
conservation of $P_\alpha$, $N$ and $Q$ to establish the existence of
a branch of solitary waves, following a method that has been used to
construct solitons of the 1-dimensional Heisenberg
ferromagnet\cite{tjonwright}.

Consider the spin-configurations that minimize the energy $E$ for
given values of $P_\alpha$, $N$, and $Q$.  By combining the
variational equations with Eq.~(\ref{eq:ll}) one finds that these 
spin-configurations have the time evolution
\begin{equation}
\frac{\partial n_i}{\partial t} = -v_\alpha \nabla_\alpha n_i - \omega
\epsilon_{ijk} n_j \hat{z}_k ,
\label{eq:time}
\end{equation}
where $v_\alpha$ and $\omega$ are Lagrange multipliers introduced to
enforce the constaints on $P_\alpha$ and $N$.  This equation shows
that the extremal spin-configurations are travelling waves, that
translate at a uniform velocity $v_\alpha$ while the magnetization
precesses around the $\hat{\bbox{z}}$ axis at an angular frequency
$\omega$.  Defining an ``energy dispersion'' $E_Q^*(P_\alpha,N)$ by
the minimum energy at given $P_\alpha$ $N$ and $Q$, the variational
equations further require
\begin{equation}
\label{eq:vel}
v_\alpha = \left.\frac{\partial E_Q^*}{\partial P_\alpha}\right|_N \;
;\;\;\;\; \omega = \left.-\frac{1}{\hbar}\frac{\partial
E_Q^*}{\partial N}\right|_{P_\alpha}.
\end{equation}

Whilst one can imagine searching for such travelling waves within each
topological subspace, note that the momentum (\ref{eq:mtm}) transforms
as $P_\alpha\rightarrow P_\alpha-Q R_\alpha$ under
$n_i(r_\alpha)\rightarrow n_i(r_\alpha+R_\alpha)$. Consequently, for a
ferromagnet that is translationally-invariant, $E_{Q\neq
0}^*(P,N)=E_{Q\neq 0}^*(0,N)$ and all modes have vanishing velocity
(\ref{eq:vel}).  For the $Q=0$ sector, however, one {\it does} expect
the energy of a translationally-invariant ferromagnet to depend on the
momentum.  In the following, we will study the properties of a
translationally-invariant ferromagnet within this $Q=0$ subspace
(henceforth we drop the subscript on $E^*_Q$).  Amongst the extremal
spin-configurations that determine $E^*(P_\alpha,N)$, we will find
configurations that are localized in space and have non-zero
$v_\alpha$ and $\omega$.  These are therefore {\it
uniformly-propagating solitary waves}\cite{rajaraman} of the
ferromagnet.

Specifically, we study an isotropic ferromagnet, for which the energy
functional for long-wavelength distortions is the $O(3)$ non-linear
$\sigma$-model
\begin{equation}
\label{eq:nls}
E^{nl\sigma}[n_i, \nabla_\alpha n_j] \equiv \frac{1}{2} \rho_s \int
d^2\bbox r \; (\nabla_\alpha n_i)^2,
\end{equation}
characterized by the spin-stiffness $\rho_s$.  Our results also apply
to extensions of this model involving additional terms that are functions of 
$Q$, $N$ and $P_\alpha$ 
(these simply shift the velocities and precession
frequencies of the modes we construct).  In the following, we will
work with units such that $\rho_s=\rho J=\hbar=1$.

We have investigated the energy dispersion $E^*(P_\alpha,N)$ of the
non-linear $\sigma$-model numerically, using a lattice discretization
of the continuum ferromagnet (details will be presented
elsewhere\cite{bigpaper}).  The results we present are for 
a square region of $249\times 249$ spins, bounded by spins fixed to
$n_z=1$.  The parameters are chosen such that the results well
represent the continuum theory (\ref{eq:nls}) defined within a
square region (of side $L$, say).  The scale-invariance of such a
theory requires that $E^*(P,N,L)=E^*(P\lambda,N\lambda^2,L\lambda)$
for all scale-factors $\lambda$, so the energy can depend only on the
``scaled momentum'' $p\equiv P/\sqrt{N}$, and the
``boundary parameter'' $\eta\equiv N/L^2$.  The properties of an
infinite system are determined by considering the limit
$\eta\rightarrow 0$.

\begin{figure}
\inseps{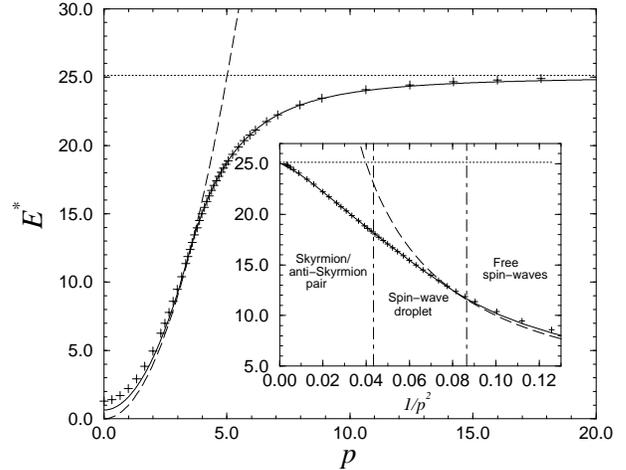}{0.46}

\caption{The energy dispersion of the isotropic ferromagnet, for
$\eta=0.032$ (solid line) and $\eta=0.064$ (crosses). The energies,
for an infinite system, of free spin-waves (dashed line) and a
non-interacting Skyrmion/anti-Skyrmion pair (dotted line) are shown
for comparison.  Inset: the same quantities as functions of $1/p^2$,
indicating the three regimes of the $\eta\rightarrow 0$ limit.}
\label{fig:energy}
\end{figure}

Figure~\ref{fig:energy} shows results for the minimum energy as a
function of $p$ at two values of the boundary parameter $\eta$.
Although weak dependences on $\eta$ do remain, it is clear that
the energy dispersion $E^*(p,\eta)$ is tending to a smooth function as
$\eta\rightarrow 0$.  The limiting configurations will describe
travelling waves with non-zero velocity
and precession frequency (\ref{eq:vel}).

\begin{figure}
\inseps{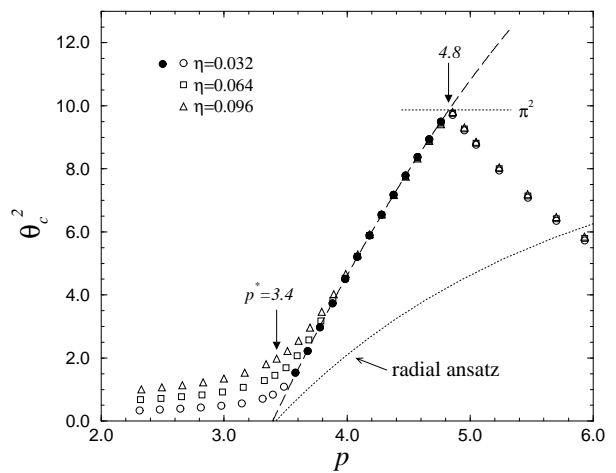}{0.46}
\caption{Variation of the square of the polar angle at the centre of
the system with scaled momentum.  An extrapolation, using the filled
circles, suggests that as $\eta\rightarrow 0$, $\theta_c$ would vanish
for $p<3.4$.  Also shown are the results for an ansatz described in
the text.}
\label{fig:t0}
\end{figure}
A careful study of the spin-configurations shows  that for
$\eta\rightarrow 0$, there is a transition from spatially-extended
configurations at small $p$, to spatially-localized configurations at
large $p$.  This is illustrated in Figure~\ref{fig:t0}, which
shows the (square of the) polar angle at the centre of the system
$\theta_c$ as a function of $p$ for several values of $\eta$.  For
$p\gtrsim 3.5$, $\theta_c$ is weakly-dependent on $\eta$, and can be
assumed to represent well an unbounded system. For $p\lesssim 3.5$,
$\theta_c^2$ decreases
steadily down to the smallest values of $\eta$ studied, showing
that the configurations remain sensitive to the presence of the
boundaries.  An
extrapolation indicates that $\theta_c^2$ would vanish at a scaled 
momentum of $p^*=3.4$ (the dashed line in Fig.~\ref{fig:t0}).

Within the assumption that $\theta_c$ does vanish as $\eta\rightarrow
0$ for $p<p^*$, the properties in this regime can be understood within
a {\it linearized} continuum theory, in which all quantities are
expanded to second-order in the polar angle $\theta(\bbox{r})$.  The
linear theory has extremal solutions
$\theta(\bbox{r})=\theta_0\cos(\pi x/L)\cos(\pi y/L)$, $\phi(\bbox{r})
=k y$, which lead to an energy dispersion $E_{SW}(p,\eta) = p^2 +
2\pi^2\eta$.  The energies of these configurations coincide with the
energy of $N$ non-interacting spin-waves confined
to a square of side $L$. We refer to them as ``free spin-wave''
states.  The numerical results for $p<p^*$ are consistent with the
minimal-energy spin-configurations of the infinite system being free
spin-wave states: the spin-configurations ressemble those of the
linearized continuum theory, and their energies approach the free
spin-wave energies of an infinite system continuously as
$\eta\rightarrow 0$ (see Fig.~\ref{fig:energy}).

In contrast, the spin-configurations above the transition, $p>p^*$,
cannot be described within the linearized theory. In this regime,
$\theta_c$ tends to a finite value as $\eta\rightarrow 0$, and the
spin-configurations are spatially-localized. These configurations
constitute the branch of propagating solitary waves that forms the
main subject of the present paper.
\begin{figure}
\inseps{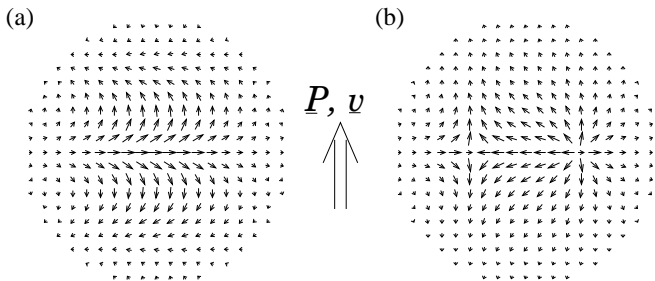}{0.69}
\vskip0.5cm
\caption{Spin-configurations of the solitary waves at (a) $p=3.64$,
and (b) $p=12.4$.  The vectors show the projection of the spin onto
the $xy$-plane for those spins with $n_z<0.98$ (for clarity only
one spin in six is plotted). The direction of the momentum and
velocity of the solitary waves is indicated. }
\label{fig:images}
\end{figure}
Typical spin-configurations of these solitary waves are shown in
Fig.~\ref{fig:images}.  For small momenta (yet larger than $p^*$) the
configurations ressemble the free-spin-wave states described above,
but are localized in space [Fig.~\ref{fig:images}(a)]. We refer to
these configurations as ``spin-wave droplets''.  For large momenta,
two vortices of opposite sign appear in the projection of the spins on
the $xy$ plane [Fig.~\ref{fig:images}(b)]. In fact, such
spin-configurations contain no singularity, since the vortices occur
around points at which $n_z=-1$, as occurs in (anti-)Skyrmion
configurations\cite{belavinpolyakov}.  Accordingly, we refer to these
configurations as ``Skyrmion/anti-Skyrmion pairs''.  The transition
between spin-wave droplets and Skyrmion/anti-Skyrmion pairs is found
to occur at $p=4.8$ (see Fig.~\ref{fig:t0}).
Note that this transition is effected by {\it continuous}
deformations of the spin-configuration.
The spin-wave droplet and Skyrmion/anti-Skyrmion pair
configurations therefore form part of the {\it same}
branch of solitary waves; the differing names we have assigned
indicate only the differences in the forms of the spin-configuration.

As the scaled momentum increases beyond $p=4.8$, the separation
between the Skyrmion and anti-Skyrmion grows continuously and the
energy approaches the energy, $8\pi$, of a non-interacting
Skyrmion/anti-Skyrmion pair\cite{belavinpolyakov}.  Asymptotic forms
that have been proposed for a widely-separated Skyrmion/anti-Skyrmion
pair\cite{foerstersilvestrov} have a number of spin-reversals $N$ that
diverges logarithmically with system size, and so correspond to the
scaled momentum $p=0$.  However, deformed versions of these
configurations can be expected to lead to asymptotic interactions of
the form $\sim 1/[p^2\log(p)]$\cite{bigpaper}.  We find an
approximately $1/p^2$ dependence (see the inset to
Fig.~\ref{fig:energy}). Logarithmic
corrections cannot be studied, 
owing to finite-size effects that appear once the
Skyrmion/anti-Skyrmion separation $\sim P/4\pi$
becomes comparable to the system size $L$ (i.e. when
$p\gtrsim 4\pi/\sqrt{\eta}$).

Previous numerical studies\cite{papzak} of the time-development of
a spin-configuration ressembling Fig.~\ref{fig:images}(b) under a
Landau-Lifshitz dynamics showed evidence for a uniform drift
parallel to the momentum. The present work establishes that solitary
waves of this form exist for the non-linear $\sigma$-model, and
determines their structures and velocities.

Having described the properties of the free spin-waves and solitary
waves, we return briefly to discuss the transition at $p^*=3.4$ that
was inferred by the extrapolation shown in Fig.~\ref{fig:t0}.
Further evidence that a transition occurs is obtained by making a
radially-symmetric ansatz for the spin-wave droplet states:
$\theta(\bbox{r})=f(|\bbox{r}|)$, $\phi =ky$. The non-linear equation
for $f(r)$ that arises from extremizing the energy has been studied in
another context\cite{kovalev}.  In the present context, the results
indicate that localized solutions are lost below $p^*_{rad}=3.4$ (see
Fig.~\ref{fig:t0}). The existence of this transition
lends weight to the idea that a transition occurs in the
unrestricted case.  Furthermore, the coincidence of the two transition
points, $p^*=p^*_{rad}$, suggests that the configuration at the
transition for the unrestricted problem is of the form assumed in the
radially-symmetric ansatz.

We now turn to discuss the application of these results to quantum
Hall ferromagnets (QHFs). These are the incompressible states
responsible for the quantized Hall effect of two-dimensional electron
systems at certain filling fractions\cite{leekane,sondhi}.  The
long-wavelength dynamics of QHFs are described by the Landau-Lifshitz
equation\cite{sondhi,stoneskyrmion}, with the Zeeman energy, $(g\mu_B
B)N$, and the non-linear $\sigma$-model term
(\ref{eq:nls}) as the leading terms in a gradient
expansion of the energy functional\cite{units}.
We will adopt this level of approximation, allowing the results presented
above to be applied directly. Higher-order terms in the gradient
expansion (e.g. long-range Coulomb forces\cite{sondhi}) will not affect the
qualitative discussion\cite{bigpaper}.

Our results provide a semiclassical description of the neutral
excitations of a QHF involving $N$ spin-reversals with total momentum
$\bbox{P}$ (this description is accurate in the limit $N\gg 1$).  The
nature of the low-energy excitation depends on the value of the scaled
momentum $p$: for $p<p^*$ it is a collection of $N$ spin-waves, each
of momentum $\bbox{P}/N$; for $p>p^*$ it is a solitary wave,
ressembling a spin-wave droplet or Skyrmion/anti-Skyrmion pair.  These
solitary waves are highly reminiscent of the ``magnetic exciton''
picture of a single spin-wave\cite{magex}: two oppositely-charged
particles move at a constant velocity perpendicular to the line
joining their centres.  Indeed, just as Skyrmions are the large-spin
versions of the quasiparticles of a QHF, these dynamical solitary
waves are the large-spin analogues of the spin-waves.

Quantum effects, not accessible within the semiclassical description,
could lead to certain differences.  If the quasiparticle gap
is less than the Skyrmion gap for vanishing Zeeman
energy\cite{wusondhisky}, the lowest-energy states at large scaled
momentum will not be Skyrmion/anti-Skyrmion pairs; they will be
widely-separated quasielectron/quasihole pairs. Also, if two
spin-waves are weakly-bound at any non-zero total
momentum\cite{wortis}, then configurations at all values of $p$ will
be spin-wave droplets. The transition found at $p=p^*$ will then be a
crossover, becoming a true transition only in the limit $N\rightarrow
\infty$.

Experimentally,
one way in which the solitary waves can be accessed is
through the application of an electric field ${\cal E}_\alpha$ to the
QHF.  For weak electric fields a quantized Hall current will flow with
exponentially-small dissipation (we assume the limit of vanishing
temperature). However, there exists a threshold value of the electric
field\cite{rasolt}, ${\cal E}_T\equiv \frac{1}{e\ell}\sqrt{16\pi
\rho_s g\mu_B B}$, above which the system can reduce its energy by
creating single spin-waves of small momentum.  The spontaneous
creation of spin-waves is expected to initiate the breakdown of the
quantum Hall effect\cite{rasolt}.
Our results indicate that, as the number of spin-waves increases, one
can expect spin-wave droplets and Skyrmion/anti-Skyrmion pairs to
form.  The appearance of Skyrmions was anticipated in a footnote to
Ref.\onlinecite{rasolt}. Our results provide a detailed description of
the non-linear structures that can develop.

Firstly, we can determine the number of spin-waves involved in the
formation of a spin-wave droplet.  For a QHF placed in a field equal
to the threshold field ${\cal E}_T$ (such that single spin-waves of
momentum $\sqrt{g\mu_B B/4\pi\rho_s}\hbar/\ell$ are formed), it is
energetically favourable for spin-waves to collapse to a spin-wave
droplet when their number reaches
$N_c={p^*}^2\rho_s/(g\mu_BB)$.

Secondly, our results show that the spin-wave droplet that forms at
this point is unstable to increasing its momentum (since the
solitary-wave group velocity is maximal at $p=p^*$).  If energy can be
lost from the electron gas to the lattice, the droplet will be torn
apart in the electric field, forming a Skyrmion/anti-Skyrmion pair of
increasing separation (these charges will eventually reach the voltage
contacts).  By the continual nucleation of spin-wave droplets, and
their relaxation to widely-separated Skyrmion/anti-Skyrmion pairs, a
d.c. current can be carried along the electric field. The quantized
Hall effect is thereby destroyed, much as superfluid flow is destroyed
by the nucleation and growth of vortex rings\cite{langerreppy} (note,
however, that for ${\cal E} \geq {\cal E}_T$ the spin-wave droplet can
nucleate without thermal fluctuations).

This breakdown mechanism could most effectively be investigated in
GaAs samples under hydrostatic pressure, for which the $g$-factor, and
thus ${\cal E}_T$, can be made to vanish\cite{maude}.  In realistic
systems, disorder will affect the breakdown mechanism strongly.  Most
importantly, the localization of the Skyrmion and anti-Skyrmion will
prevent the run-away separation expected for a clean system, and
charge transport will proceed by the condensation of spin-wave
droplets in the presence of pinned Skyrmion/anti-Skyrmion pairs.

\vskip0.2cm

I am grateful to Fabio Pistolesi, Tim Ziman and John Chalker for
helpful discussions and advice.

\widetext


\begin{thebibliography}{10}
\bibitem{units}
We reintroduce units for the QHFs: $g$ is the electron $g$-factor,
$\mu_B$ the Bohr magneton, and $\ell\equiv \sqrt{\hbar/eB}$ the magnetic
length in a field $B$.
\end{thebibliography}
\end{document}